\documentclass[reprint,twocolumn,final,aps,prb,showpacs,superscriptaddress,amsmath,amssymb,amsfonts,floatfix]{revtex4-2}

\usepackage{amssymb}
\usepackage{graphicx}
\usepackage{bm}
\usepackage{color}
\usepackage{mathrsfs}
\usepackage{textcomp, gensymb}
\usepackage{upgreek}
\usepackage[colorlinks=true,linkcolor=blue,citecolor=blue]{hyperref}

\usepackage{stackrel}
\usepackage{floatrow}

\usepackage{multirow}
\usepackage{url}
\usepackage{tabularx}
\usepackage{sidecap}
\usepackage{soul}
\usepackage{float}
\usepackage[normalem]{ulem}
\usepackage{orcidlink}
\begin{document}

\title{Electrical Modulation and Probing of Antiferromagnetism in Hybrid Multiferroic Heterostructures}

\author{Yuhan Liang}
\thanks{These authors contributed equally.}
\affiliation{School of Materials Science and Engineering, Tsinghua University, Beijing, 100084, People’s Republic of China}
\affiliation{School of Integrated Circuits and Beijing National Research Center for Information Science and Technology (BNRist), Tsinghua University, Beijing 100084, China}

\author{Huiping Han}%
\thanks{These authors contributed equally.}
\affiliation{Faculty of Materials Science and Engineering, Kunming University of Science and Technology, Kunming, Yunnan 650093, People’s Republic of China}
 
 \author{Hetian Chen}
 \thanks{These authors contributed equally.}
\affiliation{School of Materials Science and Engineering, Tsinghua University, Beijing, 100084, People’s Republic of China}

\author{Yujun Zhang}
\affiliation{Institute of High Energy Physics, Chinese Academy of Sciences, Beijing 100049, People’s Republic of China}

\author{Yi Zhang}
\affiliation{Faculty of Materials Science and Engineering, Kunming University of Science and Technology, Kunming, Yunnan 650093, People’s Republic of China}

\author{Chao Li}
\affiliation{School of Materials Science and Engineering, Tsinghua University, Beijing, 100084, People’s Republic of China}
 
\author{Shun Lan}
\affiliation{School of Materials Science and Engineering, Tsinghua University, Beijing, 100084, People’s Republic of China}

\author{Fangyuan Zhu}
\affiliation{Shanghai Synchrotron Radiation Facility, Shanghai Advanced Research Institute, Chinese Academy of Sciences, Shanghai 201204, People’s Republic of China}

\author{Ji Ma}
\affiliation{Faculty of Materials Science and Engineering, Kunming University of Science and Technology, Kunming, Yunnan 650093, People’s Republic of China}

\author{Di Yi}
\affiliation{School of Materials Science and Engineering, Tsinghua University, Beijing, 100084, People’s Republic of China}
	
\author{Jing Ma}
\affiliation{School of Materials Science and Engineering, Tsinghua University, Beijing, 100084, People’s Republic of China}

\author{Liang Wu\orcidlink{0000-0003-1030-6997}}%
\email{liangwu@kust.edu.cn}
\affiliation{Faculty of Materials Science and Engineering, Kunming University of Science and Technology, Kunming, Yunnan 650093, People’s Republic of China}

\author{Tianxiang Nan}
\email{nantianxiang@mail.tsinghua.edu.cn}
\affiliation{School of Integrated Circuits and Beijing National Research Center for Information Science and Technology (BNRist), Tsinghua University, Beijing 100084, China}

\author{Yuan-Hua Lin}
\email{linyh@tsinghua.edu.cn}
\affiliation{School of Materials Science and Engineering, Tsinghua University, Beijing, 100084, People’s Republic of China}%

\date{\today}

\begin{abstract}

The unique features of ultrafast spin dynamics and the absence of macroscopic magnetization in antiferromagnetic (AFM) materials provide a distinct route towards high-speed magnetic storage devices with low energy consumption and high integration density. However, these advantages also introduce challenges in probing and controlling AFM order, thereby restricting their practical applications. In this study, we demonstrate an all-electric control and probing of the AFM order in heavy metal (HM)/AFM insulator (AFMI) heterostructures on a ferroelectric substrate at room temperature (RT). The AFM order was detected by the anomalous Hall effect (AHE) and manipulated by the ferroelectric field effect as well as the piezoelectric effect in heterostructures of
Pt/NiO/0.7Pb(Mg$_{1/3}$Nb$_{2/3}$)O$_{3}$--0.3PbTiO$_{3}$ (PMN--PT). The non-volatile control of AFM order gives rise to a 33\% modulation of AHE, which is further evidenced by synchrotron-based X-ray magnetic linear dichroism (XMLD). Combined with the $in$-$situ$ piezoelectric response of AHE, we demonstrate that ferroelectric polarization contributes mainly to the control of the AFM order. Our results are expected to have broader implications for efficient spintronic devices.

\end{abstract}

\maketitle
\section{Introduction}
Antiferromagnets (AFM) possess distinctive properties, such as ultrafast spin dynamics and the absence of macroscopic magnetization, making them promising candidates for the development of future spintronic devices  \cite{RevModPhys.90.015005,han2023coherent,yan2019piezoelectric,chai2024voltage,liang2024observation}. Using the interaction between AFM order and itinerant electrons to detect AFM structures has attracted extensive attention  \cite{nakatsuji2015large,nayak2016large,zhou2020current,cheng2020electrical}. A non-trivial AFM order, such as noncollinear AFM or altermagnetism, can break the time-reversal symmetry (TRS) and generate intrinsic Berry curvature, resulting in a transverse Hall field to deflect spin-polarized electrons and yielding the AHE  \cite{Chen2014, nagaosa2010anomalous,nayak2016large,feng2022anomalous, PRX1, PRX2, Zhou2025,han2023coherent}. 
Thus, detecting AHE in AFM materials indicates the presence of a non-trivial AFM order, which can be manipulated by the electric field effect  \cite{wang2015electrical,huang2018electrical}, piezoelectric strain  \cite{liu2019antiferromagnetic,chen2019electric} or spin current  \cite{higo2022perpendicular,tsai2020electrical}. 
As numerous bulk AFM insulators (AFMI) host collinear AFM order (rather than altermagnetism) that protects TRS  \cite{RevModPhys.90.015005}, most studies concentrate on metallic materials with non-trivial AFM order for magnetic memories  \cite{chen2023octupole,qin2023room,higo2022perpendicular,wadley2016electrical,chou2024large}.

Recent advances show that a significant AHE around room temperature (RT) can also be observed in heterostructures comprising of AFMI and heavy metal (HM), which can be ascribed to the subtle interaction between spin current and non-trivial AFM structures during AFM--paramagnetic (AFM--PM) transition in HM/AFMI heterostructures  \cite{Liang2023,moriyama2020giant,ji2018negative,cheng2019evidence}. The AFM--PM transition temperature ($T_\text{N}$) can be regarded as the competition between AFM exchange coupling strength ($J_\text{ex}$) and thermal fluctuation, in which the $J_\text{ex}$ is sensitive to the lattice parameter and boundary condition  \cite{altieri2009prb}. Thus, in HM/AFMI heterostructures, the AFM order and correlated Berry curvature can be modulated by piezoelectric strain or ferroelectric field effect, which can be probed by AHE.

In this work, we demonstrate the electric-field control of AFM order probed by AHE in a hybrid HM/AFMI heterostructure on the piezoelectric substrate at RT. By depositing Pt/NiO heterostructures on a piezoelectric PMN--PT substrate and applying an $in$-$situ$ electric field gating ($E_\text{gate}$), we observed a non-volatile modulation of AHE conductivity of about 33\%. Through synchrotron-based X-ray magnetic linear  dichroism (XMLD) measurements, we confirmed the non-volatile modulation of AFM order in NiO that correlated to the control AHE in HM/AFMI heterostructure. Further detailed analysis on piezoelectric strain and $in$-$situ$ electric-field gating experiments indicates that the ferroelectric polarization field effect significantly dominates the control of AFM order. Our findings may facilitate the development of spintronic devices based on AFMIs.

\begin{figure*}[th]
\includegraphics[width=.9\linewidth]{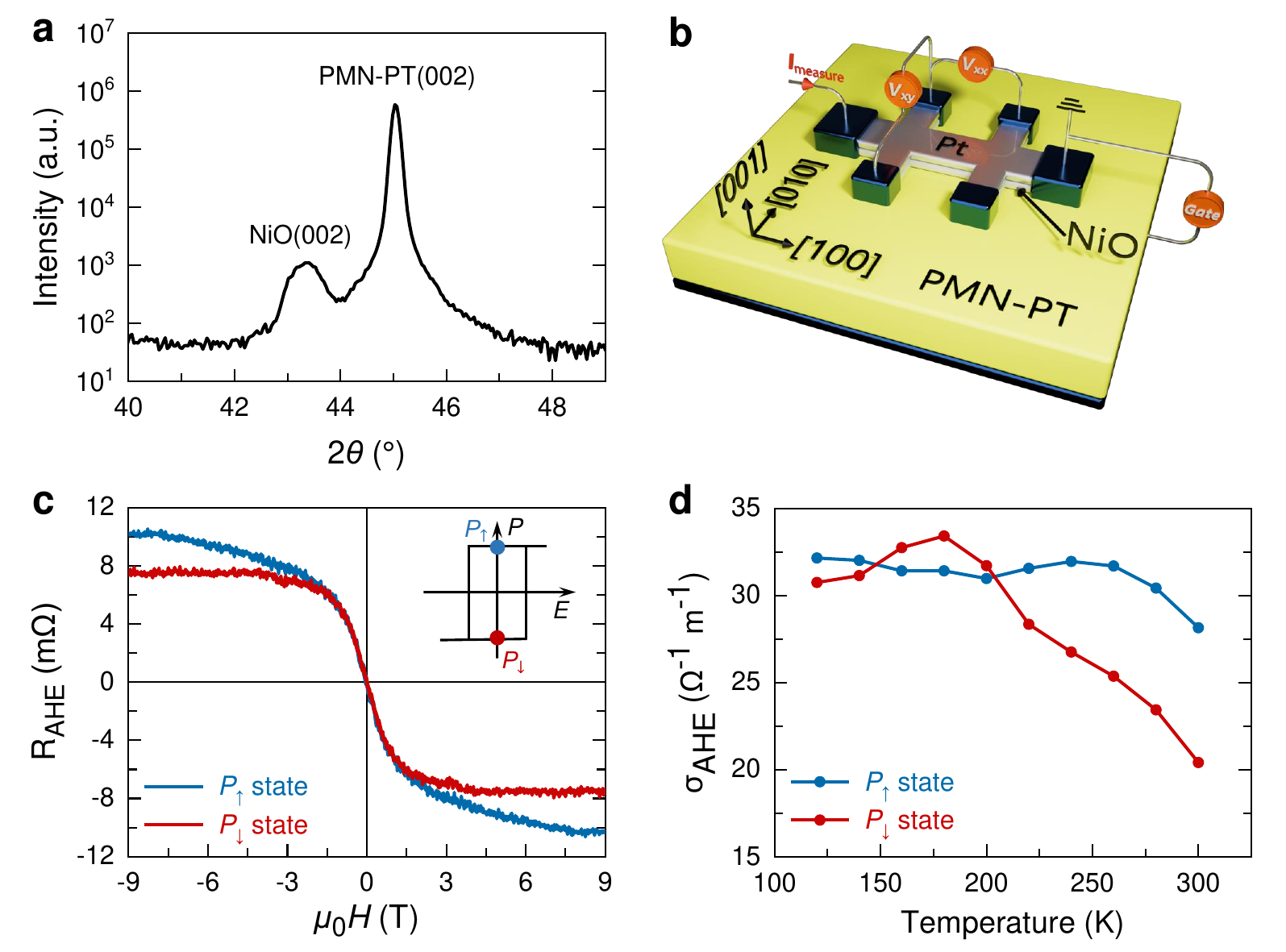}
\caption{The non-volatile electric-field control of AHE in HM/AFMI heterostructure. (a) The zoomed XRD pattern around (002) peak of 20 nm NiO thin film on PMN--PT substrate. (b) The schematic diagram of Hall bar device and transport property measurement set-up of Pt(3 nm)/NiO(3 u.c.)/PMN--PT. (c) The non-volatile electric field control of $R_\text{AHE}$ of Pt(3 nm)/NiO(3 u.c.)/PMN--PT for $P_{\uparrow}$ and $P_{\downarrow}$ states. The inset is schematic of polarization-electric field hysteresis loop. The blue and red dots represent the $P_{\uparrow}$ and $P_{\downarrow}$ states, respectively. (d) The temperature dependent $\sigma_\text{AHE}$ of Pt(3 nm)/NiO(3 u.c.)/PMN--PT for $P_{\uparrow}$ and $P_{\downarrow}$ states.}
\label{Fig1}
\end{figure*}

\section{Results and discussion}
The epitaxial NiO layer was deposited on (001)-oriented piezoelectric PMN--PT substrates by pulsed laser deposition (PLD) with a 248 nm KrF excimer laser. The NiO growth was deposited with 700 $^\circ$C substrate temperature and 4 Pa oxygen pressure. The corresponding pulse repetition was 5 Hz. After PLD deposition, the NiO/PMN--PT heterostructures were cooled to RT in an oxygen-rich atmosphere and then transferred to a magnetron sputtering chamber with a base vacuum better than $2\times10^{-8}$ Torr. The Pt was sputter deposited with an Ar pressure of 3 mTorr. The Pt/NiO/PMN--PT(001) heterostructure was fabricated into Hall bars with 20 $\mu$m width and 50 $\mu$m distance between electrode leads using standard photolithography and ion milling processes. The 100 nm Pt/5 nm Ti bilayer was then deposited through magnetron sputtering for electrode contact. The epitaxy quality is characterized by using a high-resolution X-ray diffractometer (XRD), and the thickness of films was characterized by X-ray reflectivity (XRR) (see details in \autoref{SF1} in Supplementary Materials). The sample surface morphology was characterized through atomic force microscopy. The electric field is applied through Keithley 2450 source-meter with silver paste on the PMN--PT substrate bottom as a back gating electrode. The transport property measurements were conducted in the Physical Property Measurement System (PPMS). The AHE and resistivity measurements were conducted after 20 minutes stabilization of the electric field. The ferroelectric property of PMN--PT substrate was monitored through a ferroelectric analyzer. The synchrotron-based XMLD was conducted in BL07U at Shanghai Synchrotron Radiation Facility (SSRF) with total yield electron (TEY) mode  \cite{zhu2024spatial}.

\begin{figure*}[t]
\includegraphics[width=.9\linewidth]{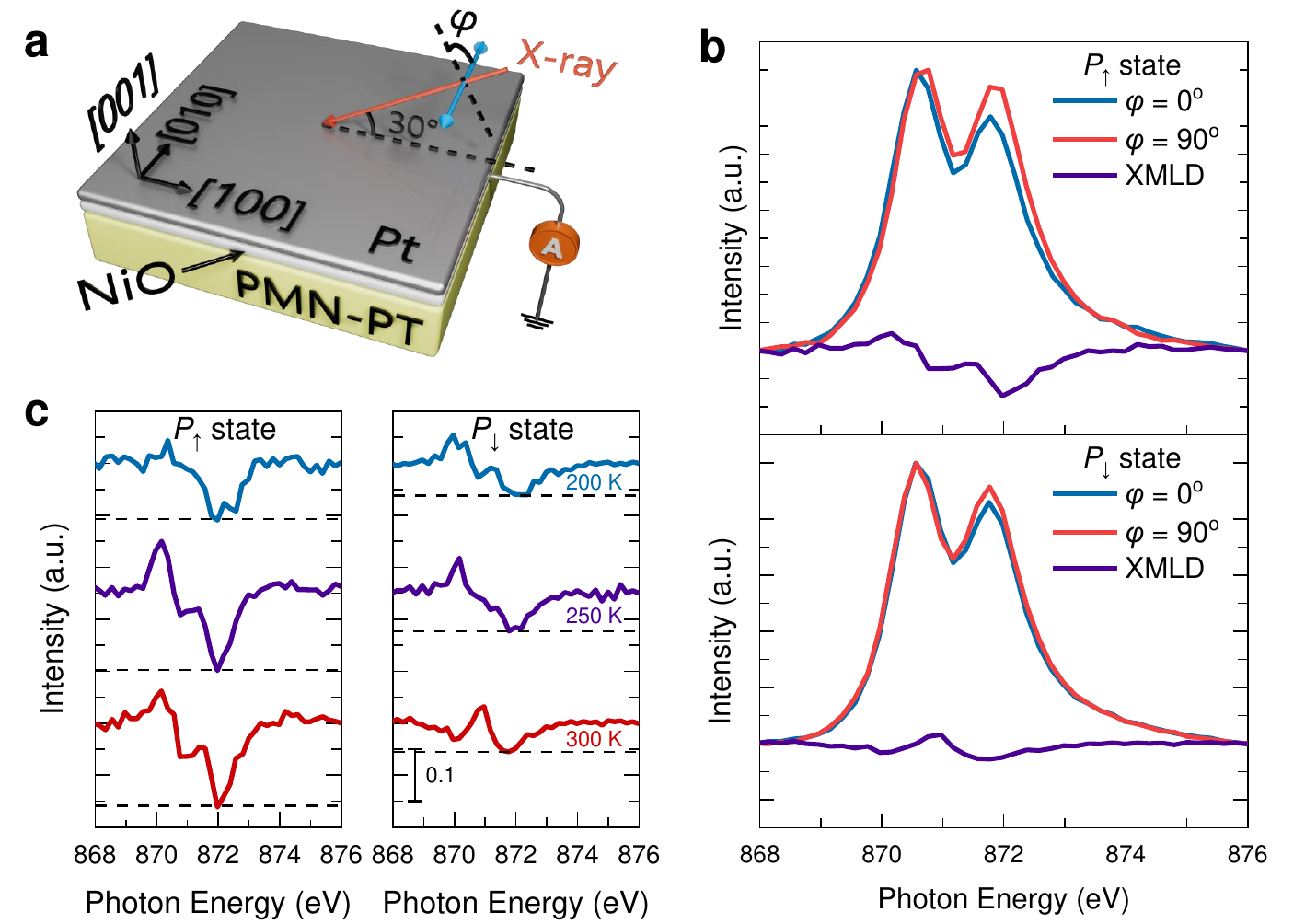}
\caption{The synchrotron-based XAS and XMLD results of Pt(3 nm)/NiO(3 u.c.)/PMN--PT heterostructure. (a) The schematic diagram of XAS measurements set-up. (b) The XAS results for $P_{\uparrow}$ and $P_{\downarrow}$ states with $\varphi=0^\circ$ and $\varphi=90^\circ$, as well as corresponding XMLD results. (c) The temperature-dependent XMLD results for $P_{\uparrow}$ state and $P_{\downarrow}$ state, respectively. The dashed line denotes the peak amplitude of XMLD signal.}
\label{Fig2}
\end{figure*}

The zoomed XRD pattern around the (002) diffraction peak of PMN--PT(001) substrate shows the high epitaxy quality of 20 nm NiO thin film, as shown in \autoref{Fig1}(a). We then decreased the thickness of NiO film and fabricated Pt(3 nm)/NiO(3 u.c.)/PMN--PT(001) heterostructure (u.c. is the abbreviation of unit cells), in which the AFM--PM transition is around RT  \cite{Liang2023}. The surface morphology of Pt(3 nm)/NiO(3 u.c.)/PMN--PT(001) heterostructure is shown in \autoref{SF2}, Supplementary Materials, revealing a roughness of about 700 pm. The AHE and resistivity measurements were then conducted in a Hall bar device with measurement current $I_\text{measure}$ of 0.5 mA applied along the in-plane [100] direction, corresponding to a current density of about 8.3$\times 10^{5} \, \text{A/cm}^{2}$ about two orders lower than the value for substaintial spin-orbit torque effects \cite{ChenxzSwitching2018,moriyama2018spin}, as schematically shown in \autoref{Fig1}(b). By applying $E_{\text{gate}}=\pm 4$ kV/cm, the PMN--PT substrate can be poled into opposite ferroelectric polarization $P$. This electric field is homogeneously out-of-plane, minimizing the possible spin accumulation induced by in-plane electric field \cite{cai2017electric}. After poling of the PMN--PT substrate, the $E_\text{gate}$ was removed and PMN--PT substrate relaxed into two remnant states with ferroelectric polarization about $\pm$20 $\mu\text{C}/\text{cm}^2$ (noted as states $P_{\uparrow}$  and $P_{\downarrow}$, respectively), see polarization-electric field hysteresis loop in \autoref{SF3}, Supplementary Materials. The resistivity variation of the device was confirmed to be less than 0.4\% between $P_{\uparrow}$  and $P_{\downarrow}$ states, as shown in \autoref{SF4}, Supplementary Materials. The AHE resistance $R_\text{AHE}$ was monitored by sweeping the out-of-plane external magnetic field $\mu_{0}H$ at RT for the states $P_{\uparrow}$ and $P_{\downarrow}$, as shown in \autoref{Fig1}(c). 
It should be noted that the Hall data is an odd function as the sweeping magnetic field, avoiding the mixing of magnetoresistance, which is an even function. The ordinary Hall effect background induced by Lorentz force has been subtracted with  coefficient about $3.7 \times 10^{-7} \,\Omega\,\text{Oe}^{-1}$ for the states $P_{\uparrow}$ and $3.6 \times 10^{-7} \,\Omega\,\text{Oe}^{-1}$ for the states $P_{\downarrow}$, respectively  (see \autoref{SF5} in Supplementary Materials for the raw Hall resistance results).
It clearly shows the non-volatile electric-field control of AHE, as the saturated $R_\text{AHE}$ for $P_{\uparrow}$ and $P_{\downarrow}$ states is 7.5~m$\Omega$ and 10~m$\Omega$, respectively. 

\begin{figure*}[t]
\includegraphics[width=.9\linewidth]{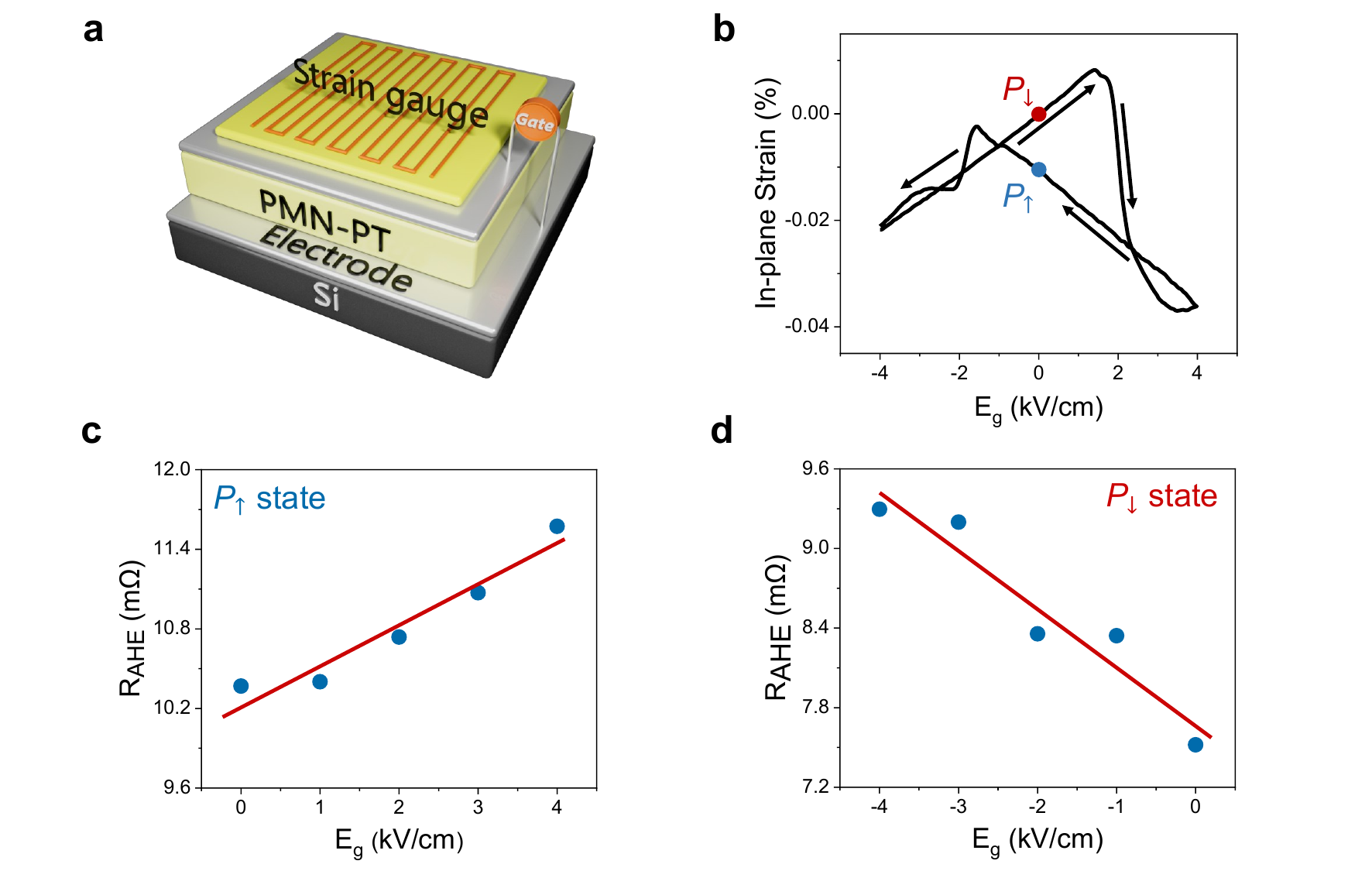}
\caption{The piezoelectric effect on the AHE of Pt(3 nm)/NiO(3 u.c.)/PMN--PT heterostructure. (a) The schematic of measuring in-plane strain by strain gauge. (b) The monitored in-plane strain by sweeping out-of-plane electric field. The arrows indicate the sweeping direction of electric field. The blue and red dots represent the $P_{\uparrow}$  and $P_{\downarrow}$ states, respectively. (c) and (d), The electric-field dependent $R_\text{AHE}$ for heterostructure with (c) $P_{\uparrow}$ and (d) $P_{\uparrow}$ states, respectively. The red lines indicate the linear fitting.}
\label{Fig3}
\end{figure*}

Considering the $R_\text{AHE}$ originates from the combination between AHE conductance $\sigma_\text{AHE}$ and longitudinal resistivity $\rho_{xx}$, we measured the $\rho_{xx}$ of two remnant states and estimated the corresponded $\sigma_\text{AHE}$ at various temperatures according to $\sigma_\text{AHE}=\rho_{\text{AHE}}/(\rho_{\text{AHE}}^2+\rho_{xx}^2)$, where $\rho_{\text{AHE}}=R_\text{AHE}t$ and $t=$ 3 nm is the thickness of the Pt layer. See detailed data on temperature-dependent AHE results in \autoref{SF6}, Supplementary Materials. The calculated $\sigma_\text{AHE}$ are shown in \autoref{Fig1}(d). At RT, the $\sigma_\text{AHE}$ for the $P_{\uparrow}$ and $P_{\downarrow}$ states is approximately 28 and 20 $\mathrm{\Omega^{-1}~m^{-1}}$, respectively, corresponding to 33\% modulation of $\sigma_\text{AHE}$ at RT. In the whole temperature range from 300 to 120 K, for $P_{\uparrow}$ state, a local maximum value of $\sigma_\text{AHE}$ approximately 32 $\mathrm{\Omega^{-1}~m^{-1}}$ is observed around 240 K, while a local maximum value of 33 $\mathrm{\Omega^{-1}~m^{-1}}$ is observed around 180 K for $P_{\downarrow}$ state. These results suggest that the AFM strength of ultrathin NiO films is effectively modulated by $E_\text{gate}$.

\begin{figure*}[t]
\includegraphics[width=1\linewidth]{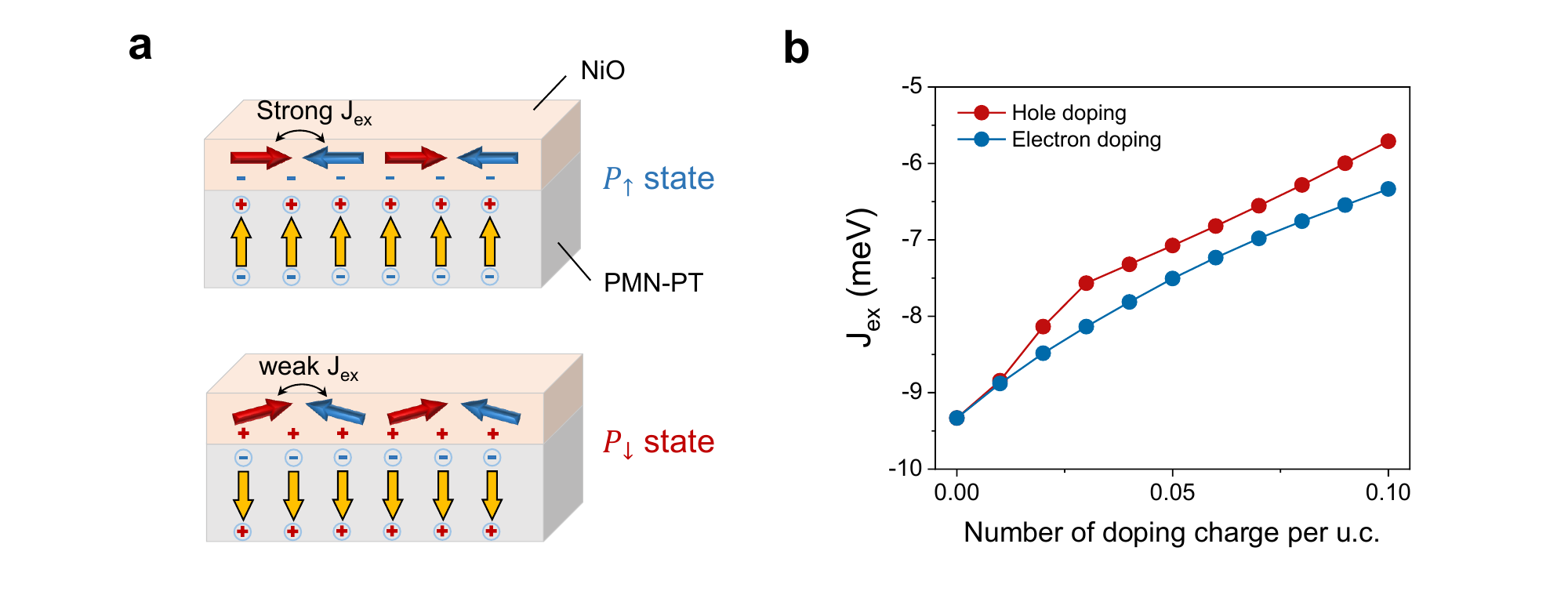}
\caption{The DFT calculation of $J_\text{ex}$ controlled by ferroelectric field effect. (a) The schematic ferroelectric field effect on $J_\text{ex}$. For $P_{\uparrow}$ state, the positive bound charge on the NiO/PMN--PT interface induce electron doping into NiO layer, which possesses strong $J_\text{ex}$. For $P_{\downarrow}$ state, negative bound charge on the NiO/PMN--PT interface induce hole doping into NiO layer, which possesses weak $J_\text{ex}$. The yellow arrows are ferroelectric polarizations. The symbols ``$+$" and ``$-$" with circles denote positive or negative bound charge, while those without circles denote hole/electron doping.  (b) the DFT calculated $J_\text{ex}$ as a function of the number of charge per u.c. ($N$) for electron or hole doping.}
\label{Fig4}
\end{figure*}

To confirm the modulation of the AFM strength of ultrathin NiO film  by $E_\text{gate}$, we utilized synchrotron-based XMLD measurements at the Ni $L_2$ edge using the same method described in Refs. \cite{alders1998prb,altieri2009prb}. The experimental set-up is schematically shown in \autoref{Fig2}(a), as the X-ray is incident in (010) plane with an angle of 30$^\circ$ with respect to the sample surface. The polarization vector $\vec{E}$ of X-ray can be modulated with angle $\varphi$ ($\varphi=0^\circ$ or $90^\circ$) to the (010) plane. The $L_{2}$ intensity ratio in the X-ray absorption spectrum (XAS) at the Ni $L_{2}$ edge, defined as the ratio of the lower to higher energy peak intensities, is affected by the alignment of $\vec{E}$ with the Ni$^{2+}$ spin easy axis  \cite{stohr1999images} and exhibits linear dichroism. The linear dichroism effect can arise from the XMLD effect due to AFM order and the crystal field effect; however, the latter can be disregarded since our observations show an insignificant peak energy shift at the $L_{3}$ edge (see \autoref{SF7} in Supplementary Materials), which also exclude the Ni defects in Pt (see \autoref{SF8} in Supplementary Materials). By varying the angle $\varphi$, the $L_{2}$ intensity ratio of the normalized XAS signal could change and show the XMLD effect, which indicates the AFM order strength  \cite{altieri2009prb}. In this work, we define the XMLD signal = XAS$(\varphi=0^\circ)-\text{XAS}(\varphi=90^\circ)$. The typical normalized XAS results with $\varphi=0^\circ$ and $\varphi=90^\circ$ and the corresponding XMLD signal for $P_{\uparrow}$ and $P_{\downarrow}$ states are shown in \autoref{Fig2}(b) and \autoref{Fig2}(c), respectively. For both states, the $L_{2}$ higher energy peak intensity always reaches its maximum value (or the minimum $L_{2}$ intensity ratio) with $\varphi=90^\circ$, indicating the Ni$^{2+}$ spin easy axis lying in the sample plane and consistent with the reported easy-plane AFM structure of NiO thin film  \cite{wu2008analysis,gray2019spin,baldrati2018full}. 

 Then we utilized the XMLD signal to reveal the AFM order strength modulated by $E_\text{gate}$. The XMLD is significant for $P_{\uparrow}$ state, while it becomes weaker for $P_{\downarrow}$ state at 300 K, see \autoref{Fig2}(b), supporting our previous demonstration of the non-volatile modulation on AFM order strength by $E_\text{gate}$ according to AHE measurements in \autoref{Fig1}. In addition, we also conducted XAS measurements with temperatures ranging from 300 K to 200 K, and obtained temperature-dependent XMLD results, as shown in \autoref{Fig2}(c). For $P_{\uparrow}$ state, the XMLD signal remains significant as temperature decreases from 300 K to 200 K. We note that the XMLD signal shows a non-monotonic trend, as it becomes relatively weaker at 200 K than at 250 K, which can be ascribed to the AFM spin reorientation at low temperature  \cite{zhu2014antiferromagnetic,wu2006tailoring}. Furthermore, the XMLD signal from the $P_{\downarrow}$ state shows a similar non-monotonic trend but remains less than the XMLD signal magnitude from the $P_{\uparrow}$ state. This trend aligns with the AHE results in \autoref{Fig1}(d), showing that the AFM order strength of $P_{\downarrow}$ state is weaker than that in $P_{\uparrow}$ state, which is also consistent with the reported correlation between AHE and AFM order strength in AFM conductors  \cite{nakatsuji2015large,liu2018electrical,nagaosa2010anomalous}.

As a relaxor ferroelectric with significant piezoelectric effect, after applying reverse $E_\text{gate}$, the observed non-volatile control of AHE could be a combination of ferroelectric polarization switching and piezoelectric strain effect. The PMN--PT has a rhombohedral crystal structure, with the ferroelectric polarization of PMN--PT along $\left< 111 \right>$ directions in pseudo-cubic notation  \cite{chen2019giant,yan2019piezoelectric}. Thus, a critical reverse $E_\text{gate}$ applied along the [001] direction can lead to 71$^\circ$, 109$^\circ$ and 180$^\circ$ switching of ferroelectric polarization  \cite{yang2014bipolar}. Although saturation $E_\text{gate}$ favors 180$^\circ$ switching and volatile in-plane strain after removing the electric field, minor 71$^\circ$ or 109$^\circ$ switching can also remain and induce non-volatile strain that modulates the AFM order  \cite{ba2021electric}. To clarify this issue, we measured the in-plane piezoelectric strain ($\varepsilon$) by using a strain gauge with continuous measurements on PMN--PT(100) substrate, see schematic diagram in \autoref{Fig3}(a). The in-plane piezoelectric strain as a sweep $E_\text{gate}$ along the [001] direction is displayed in \autoref{Fig3}(b), showing a butterfly-shaped strain hysteresis loop with an asymmetry component. By defining the piezoelectric strain coefficient $\gamma = \left|d\varepsilon/dE_{\text{gate}}\right|$, we found the $\gamma_{\uparrow} = 0.0063~\mathrm{\%~cm/kV}$ for $P_{\uparrow}$ state, and $\gamma_{\downarrow} = 0.0058~\mathrm{\%~cm/kV}$ for $P_{\downarrow}$ state, respectively. Two non-volatile states with in-plane strain about $-0.01\%$ corresponding to $P_{\uparrow}$ state and about 0\% corresponding to $P_{\downarrow}$ state were observed after removing $E_\text{gate}$. 
Next, to clarify the dominant effect in the non-volatile control of AFM order, we studied the piezoelectric response of AHE. The polarization is held upward for $P_{\uparrow}$ state and downward for $P_{\downarrow}$ state, respectively, and the $E_\text{gate}$ is applied $in$-$situ$ to the device up to $\pm 4$ kV/cm; see detailed raw data in \autoref{SF9}, Supplementary Materials. The results are shown in \autoref{Fig3}(c) and \autoref{Fig3}(d), displaying similar magnitude of magnetoelectric coupling coefficient $\beta=d\text{R}_{\text{AHE}}/dE_{\text{gate}}$ for both states. We found $\beta_{\uparrow}=0.31~\mathrm{m\Omega~cm /kV}$ for $P_{\uparrow}$ state and $\beta_{\downarrow}=0.44~\mathrm{m\Omega~cm /kV}$ for $P_{\downarrow}$ state. The increase of in-plane compressive strain gives rise to enhancement of $R_\text{AHE}$, which can be understood as the enhancement of AFM interaction $J_\text{ex}$ by reducing interatomic spacing  \cite{liu2018electrical}. However, the magnitude of $R_\text{AHE}$ of $P_{\downarrow}$ state is always lower than that of $P_{\uparrow}$ state in the range of applied $E_\text{gate}$. Considering the strain difference of about 0.01\% between $P_{\uparrow}$ and $P_{\downarrow}$, the pure modulation of $R_\text{AHE}$ originates from the remnant strain can be calculated through $\Delta R_\text{AHE}^\text{strain} = \frac{dR_\text{AHE}}{d\varepsilon}\Delta\varepsilon = \frac{dR_\text{AHE}}{dE_{\text{gate}}}\cdot\frac{dE_{\text{gate}}}{d\varepsilon}\Delta\varepsilon = \beta/\gamma\cdot\Delta\varepsilon $. Using the coefficiencies from experimental results, we estimated the $\Delta R_\text{AHE}^\text{strain}$ is about 0.63$\pm$0.14~m$\Omega$, which can only contribute to about 25\% of the non-volatile modulation of the $R_\text{AHE}$ in \autoref{Fig1}(c). Thus, in our system, the field effect imposed by ferroelectric polarization of PMN--PT mainly contributes to the non-volatile control of AFM order strength and corresponding $R_\text{AHE}$.

\vspace{0.5em}
Due to the bound charge carried by ferroelectric polarization, electron/hole doping into the ultrathin NiO layer could happen for $P_{\uparrow}$/$P_{\downarrow}$ states  \cite{gu2024ferroelectric,song2017recent,si2019ferroelectric}. In correlated magnetic materials, this doping usually results in modulation of $J_\text{ex} $ \cite{zheng2018ambipolar}, as schematically shown in \autoref{Fig4}(a). To qualitatively understand the impact of hole/electron doping on $J_\text{ex}$ in the NiO layer, we conducted theoretical calculations based on density functional theory (DFT) to investigate $J_{ex}$. The number of doping charge per unit cell ($N$) for either electron or hole doping is simply estimated by $N=PV_0/de\approx~0.07$, where $P\approx 20~ \mu\text{C}/\text{cm}^2$ is the magnitude of ferroelectric polarization, $V_0$ is the volume of the NiO unit cell, $d \approx~1.2~\text{nm}$ is the thickness of the NiO layer, and $e$ is the electron charge. Thus, we calculated $J_\text{ex}$ considering $N$ from 0 to 0.1 with electron or hole doping. The NiO remains insulating according to the band structure, see details in \autoref{SF10}, Supplementary Materials. Without doping, $J_\text{ex}$ is calculated to be about $-9.3$ meV, located in the reasonable range within previous theoretical studies  \cite{PhysRevB.66.064434,PhysRevB.65.155102}. Furthermore, our DFT calculations revealed that both hole and electron doping in the NiO layer lead to a weakening of the $J_\text{ex}$. However, this modulation depends on the type of doping and shows strong asymmetry, which has been intensively demonstrated in correlated materials  \cite{zheng2018ambipolar,parschke2017correlation,segawa2010zero}. The $J_\text{ex}$ of electron doping is always larger than hole doping, giving rise to stronger AFM strength probed by larger AHE at RT (\autoref{Fig1}), as well as a stronger XMLD signal (\autoref{Fig2}). The DFT calculation is further supported by a Pt(3 nm)/NiO(4 u.c.)/PMN-PT control sample (see \autoref{SF11} in Supplementary Materials), in which the nonvolatile modulation of AHE resistance is much smaller due to thicker NiO, aligning with the scenario of the ferroelectric field-effect.

Our results highlight that the AFM state can be effectively probed through interfacial effects using the AHE, despite the spin-independent nature of bulk AFMs that normally preserves spin degeneracy. Although demonstrated here with insulating NiO, similar phenomena could be extended to conducting AFMs through precise thickness control, opening new avenues for designing functional interface-engineered AFM devices \cite{YANG2025100142,zhao2025collinearantiferromagnetictunneljunctions}. A possible asymmetric spin–orbit torque driven AFM switching may also emerge with an appropriate crystalline orientation under a piezoelectric strain–modulated energy barrier governed by the AFM strength \cite{zhang2025deterministicswitchingneelvector}.

\section{Conclusion}
In conclusion, we report the electric-field manipulation of AFM order in hybrid HM/AFMI heterostructures on PMN--PT substrate. By applying $E_\text{gate}$, a non-volatile 33\% control of $\sigma_\text{AHE}$ is observed at RT. Using XMLD measurements, the modulation of $R_\text{AHE}$ is ascribed to the control of AFM order. In addition, by monitoring the piezoelectric effect on $R_\text{AHE}$, we conclude that the non-volatile control of $R_\text{AHE}$ is mainly contributed by ferroelectric polarization of PMN--PT, which modulates the image charge screening effect from Pt for the enhancement or reduction of $J_\text{ex}$. The DFT calculation shows the asymmetric $J_\text{ex}$ dependence on the electron or hole doping induced by the ferroelectric field effect on the NiO layer. 
Our results may facilitate the development of AFM-based spintronic devices.

\section*{ACKNOWLEDGMENTS}
We thank the staff at beamline BL07U of the Shanghai Synchrotron Radiation Facility (SSRF) for fruitful discussions and experimental assistance. L.W. acknowledges the support from the National Natural Science Foundation of China (Grant No. 52102131). Y.-H.L acknowledges the support from the Basic Science Center Project of the National Natural Science Foundation of China (Grant No. 52388201). D.Y. acknowledges the support from National Natural Science Foundation of China (Grant No. 52250418).

\section*{Data Availability}
The data that support the findings of this study are available from the corresponding author upon reasonable request.

\bibliography{ref}

\clearpage            
\onecolumngrid
\setcounter{figure}{0}
\setcounter{figure}{0}
\renewcommand{\thefigure}{S\arabic{figure}}
\renewcommand{\theHfigure}{S\arabic{figure}} 

\begin{center}
    \textbf{Supplementary Materials}
\end{center}

\begin{figure}[h]
	\includegraphics[width=0.8\linewidth]{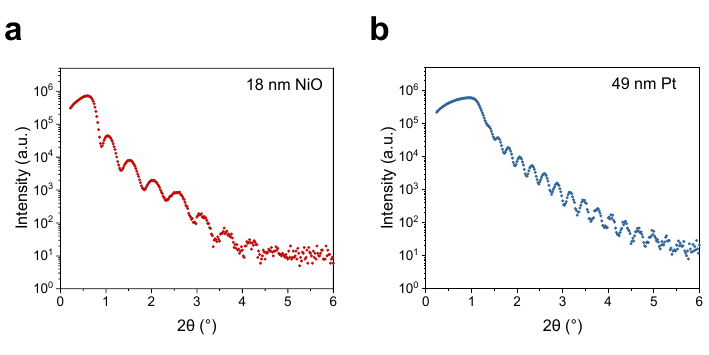}
	\caption{The XRR raw data of (a) 18 nm NiO and (b) 49 nm Pt.}
	\label{SF1}
\end{figure}

\begin{figure}[h]
	\includegraphics[width=0.45\linewidth]{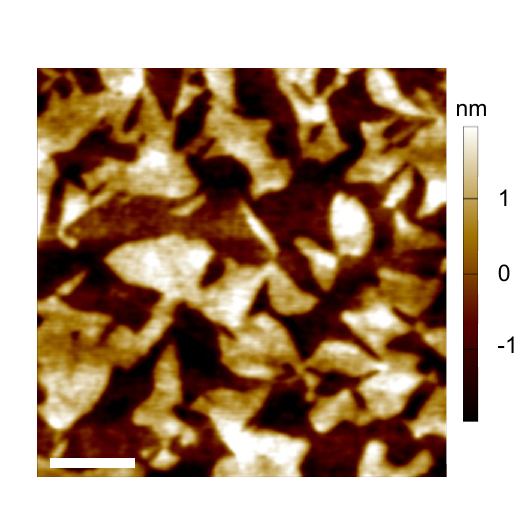}
	\caption{The atomic force microscopy image of Pt(3 nm)/NiO(3 u.c.)/PMN-PT heterostructure. The scale bar is 1 $\mu$m.}
	\label{SF2}
\end{figure}

\begin{figure}[h]
	\includegraphics[width=1\linewidth]{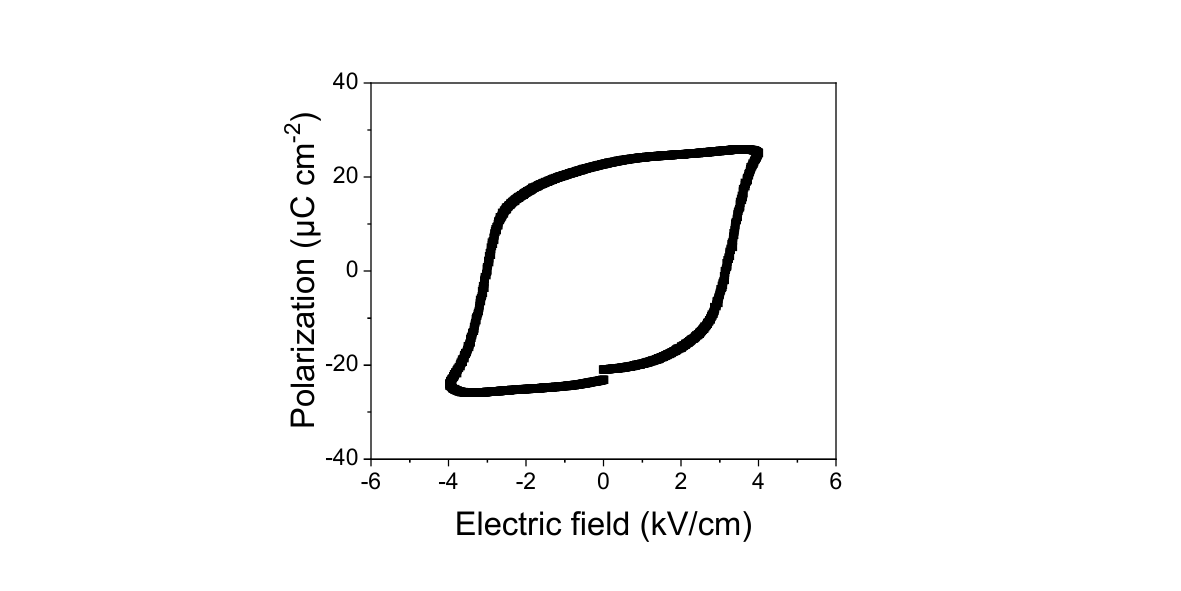}
	\caption{The polarization--electric field hysteresis loop of PMN--PT substrate.}
	\label{SF3}
\end{figure}

\begin{figure}[h]
	\includegraphics[width=1\linewidth]{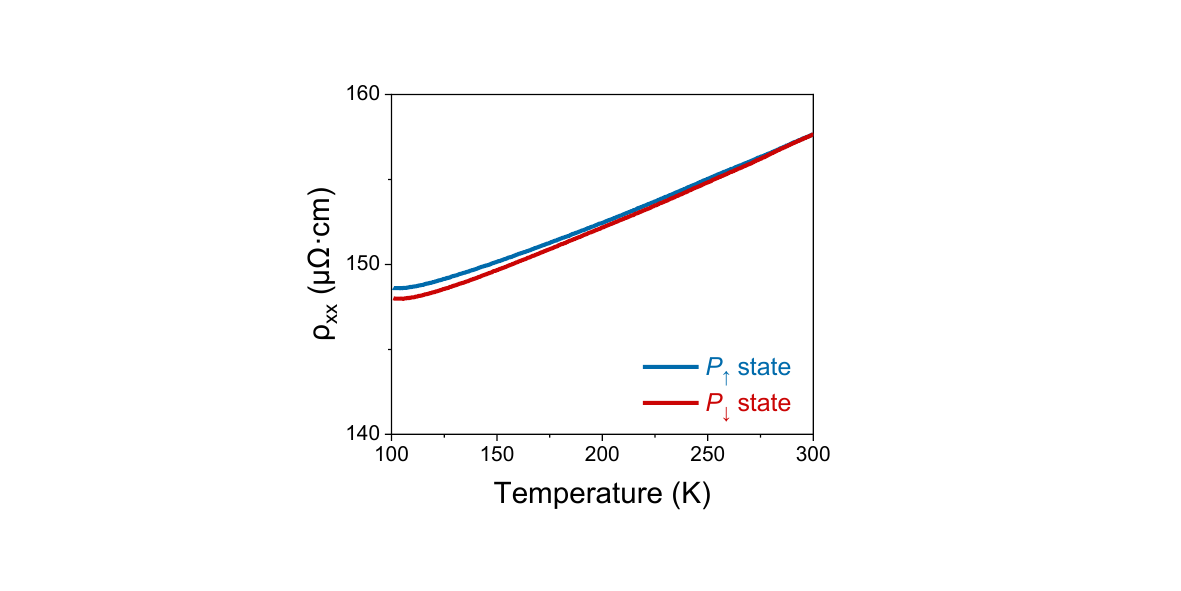}
	\caption{The temperature-dependent resistivity for $P_\uparrow$  state and $P_\downarrow$ state, respectively.}
	\label{SF4}
\end{figure}

\begin{figure}[h]
	\includegraphics[width=1\linewidth]{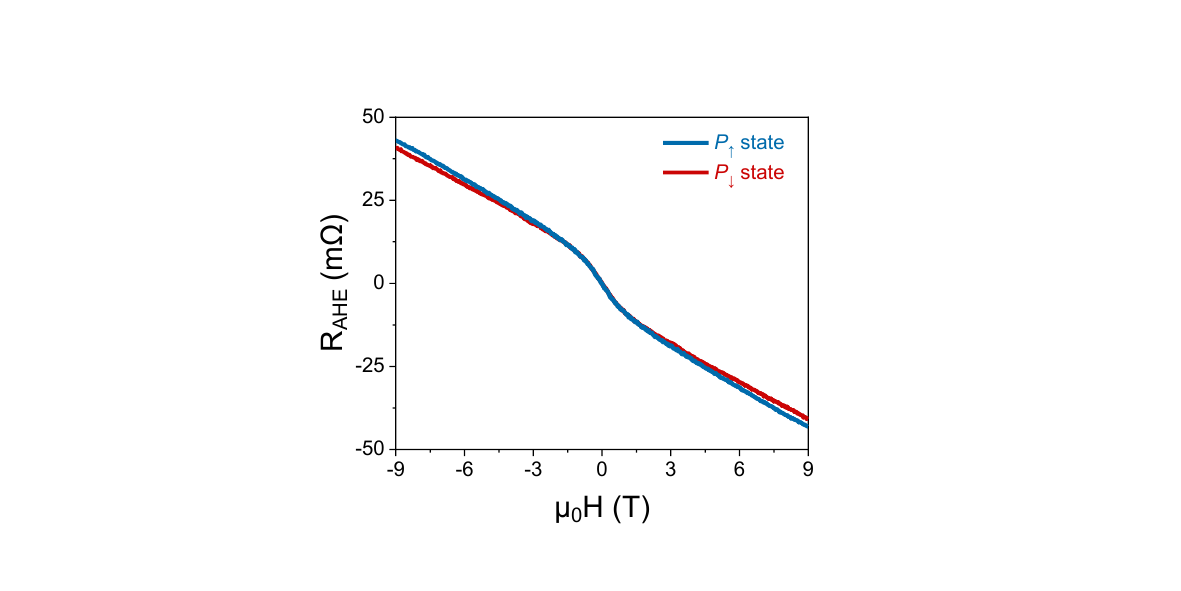}
	\caption{The raw data of AHE resistance as a function of out-of-plane magnetic field for $P_\uparrow$ state and $P_\downarrow$ state, respectively.}
	\label{SF5}
\end{figure}

\begin{figure}[h]
	\includegraphics[width=1\linewidth]{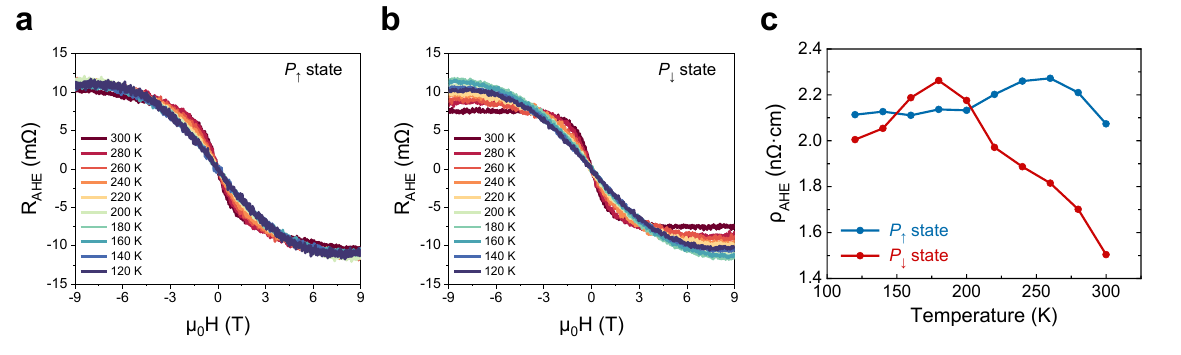}
	\caption{The AHE resistance under different temperature for (a) $P_\uparrow$ state and (b) $P_\downarrow$ state, respectively. The temperature dependent saturated $R_\text{AHE}$ for both states are summarized in (c).}
	\label{SF6}
\end{figure}

\begin{figure}[h]
	\includegraphics[width=0.8\linewidth]{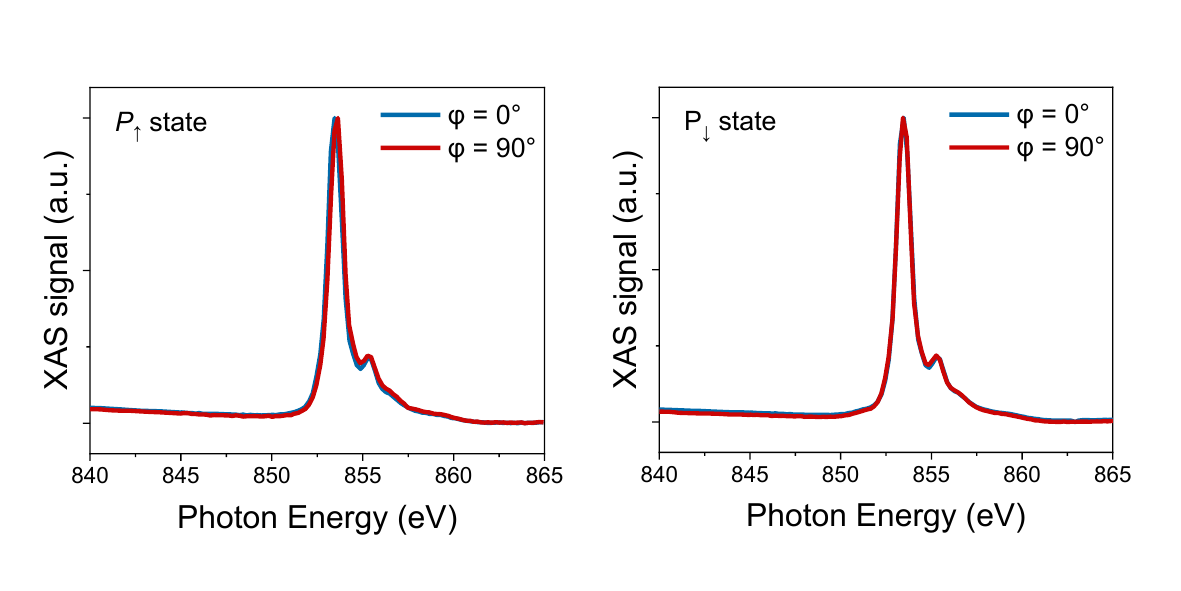}
	\caption{The Ni $L_3$ edge XAS results for $P_\uparrow$ state and $P_\downarrow$ states, respectively.}
	\label{SF7}
\end{figure}

\begin{figure}[h]
	\includegraphics[width=0.8\linewidth]{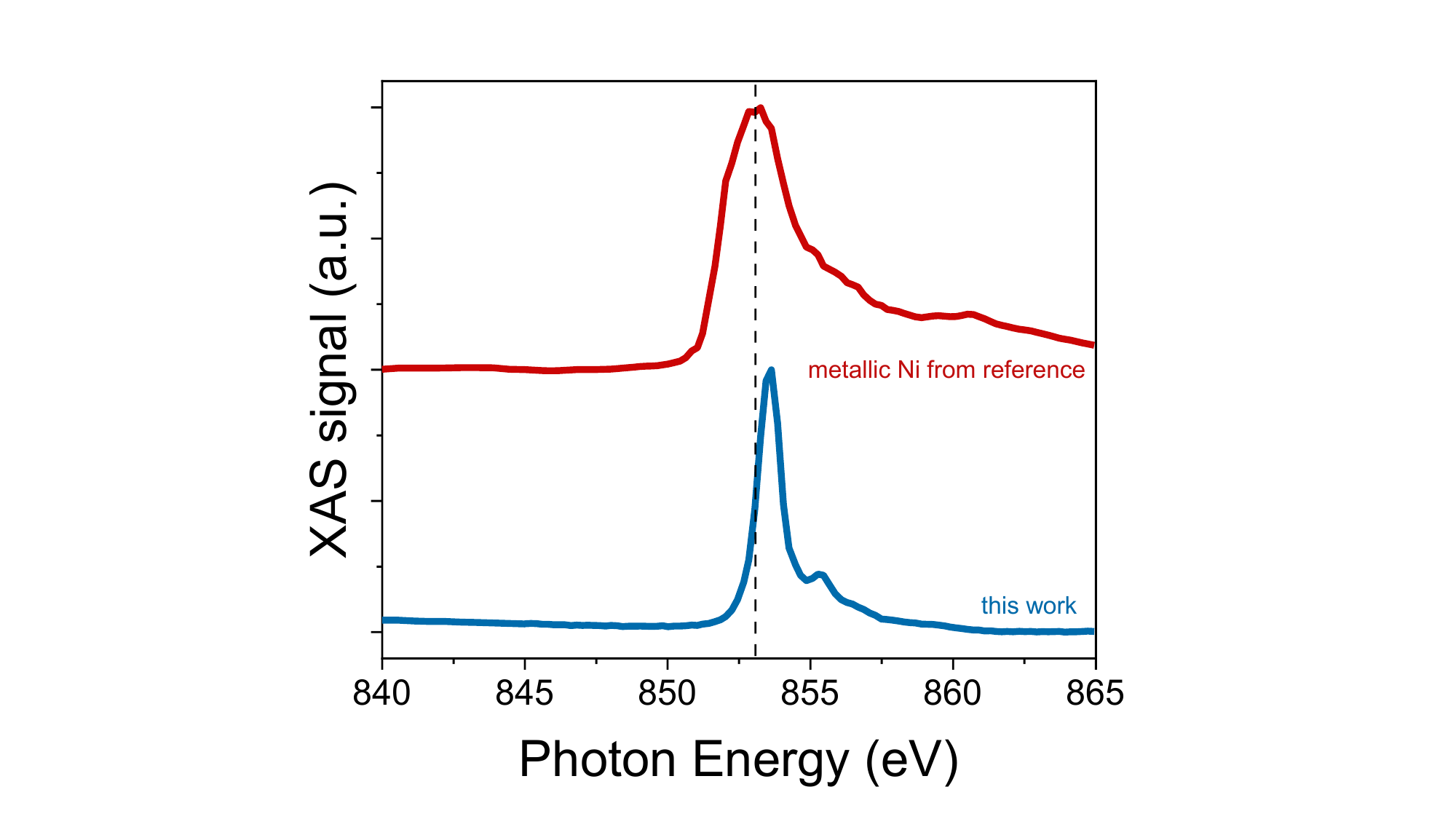}
	\caption{Ni $L_3$-edge spectra of metallic Ni [Phys. Rev. B 101, 115137 (2020)] and this work. The dash line indicates the peak position of metallic Ni, excluding the Ni defects in Pt for AHE.}
	\label{SF8}
\end{figure}

\begin{figure}[h]
	\includegraphics[width=0.8\linewidth]{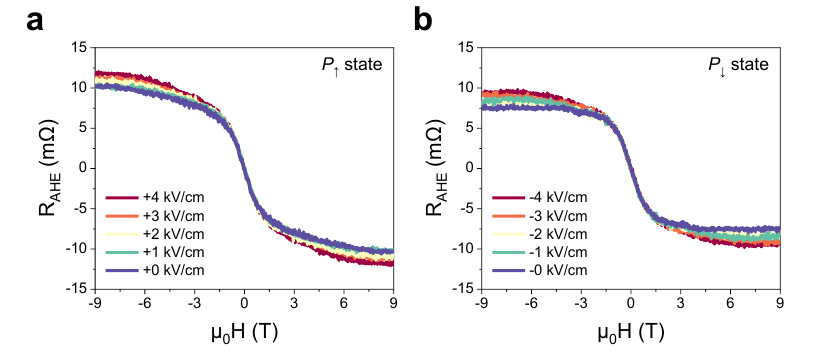}
	\caption{The AHE resistance under different E$_{\text{gate}}$ for (a) $P_\uparrow$ state and (b) $P_\downarrow$ state, respectively.}
	\label{SF9}
\end{figure}

\clearpage
\noindent
\textbf{The DFT calculation of $\bm{J}_\mathbf{ex}$ in electron/hole doped NiO}

\bigskip

The DFT calculation is performed using the Vienna \textit{Ab initio} Simulation Package (VASP). The projector-augmented-wave (PAW) pseudopotential was utilized to treat the core electrons. The Perdew–Burke–Ernzerhof (PBE) functional, based on the generalized gradient approximation (GGA) for exchange-correlation, was chosen to simulate the electron interactions. The cutoff energy is set to 450~eV for all calculations. The Brillouin zone is sampled by a Monkhorst-Pack  $4 \times 4 \times 4$  $ k $-point mesh. To simplify the calculation, we only calculated the nearest exchange interaction between Ni atoms, expressed as $J_\text{ex}$. The schematic of the NiO primary unit cell is shown in Figure~\ref{SF8}(a), as the rhombohedral unit cell vectors are given as $(1,1/2,1/2)_a$, where $a = 4.2$ \AA. For each individual Ni that couples to the nearest Ni, the exchange coupling energy is expressed as $J_\text{ex}\mathbf{S}_1\cdot\mathbf{S}_2$, where the magnitude of $\mathbf{S}_{1,2}$ equals 1 for Ni$^{2+}$. Thus, our strategy is to calculate the energy of AFM NiO ($E_\text{AFM}$ where $J_\text{ex}<0$), as well as the energy of ferromagnetic (FM) NiO ($E_\text{FM}$ where $J_\text{ex}>0$). The density of states (DOS) for AFM NiO and FM NiO are shown in Figure~\ref{SF8}(b) and Figure~\ref{SF8}(c) with a certain band gap, indicating that both are insulators. However, the band gap of FM NiO is narrower than that of AFM NiO. By determining the energy difference between AFM and FM NiO, represented by $\Delta E = E_\text{AFM}-E_\text{FM}$ and dependent on the number of charges per unit cell ($N$) as illustrated in Figure~\ref{SF8}(d), we can subsequently compute the exchange interaction $J_\text{ex}=\Delta E/(24S^2)$.

\begin{figure}[h]
	\includegraphics[width=1\linewidth]{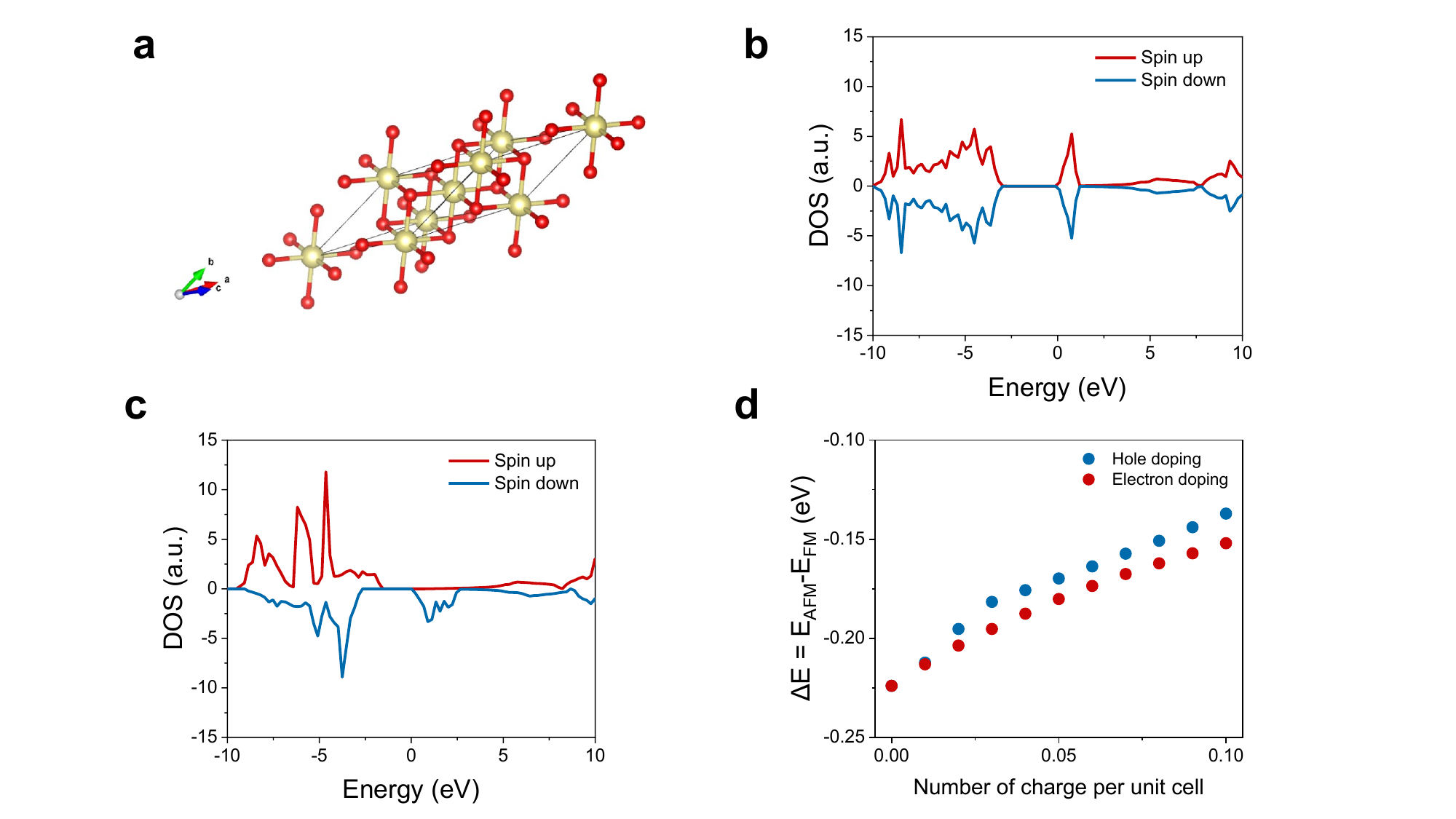}
	\caption{The details of DFT calculation. (a) The schematic of primary unit cell of NiO for DFT calculation. Only two Ni atoms are contained. The yellow balls are Ni atoms, red balls are oxygen atoms. (b) The calculated DOS as a function of energy of AFM NiO. (c) The calculated DOS as a function of energy of FM NiO. (d) The calculated $\Delta E$ as a function of number of charge per unit cell.}
	\label{SF10}
\end{figure}

\clearpage
\noindent

\begin{figure}[h]
	\includegraphics[width=0.8\linewidth]{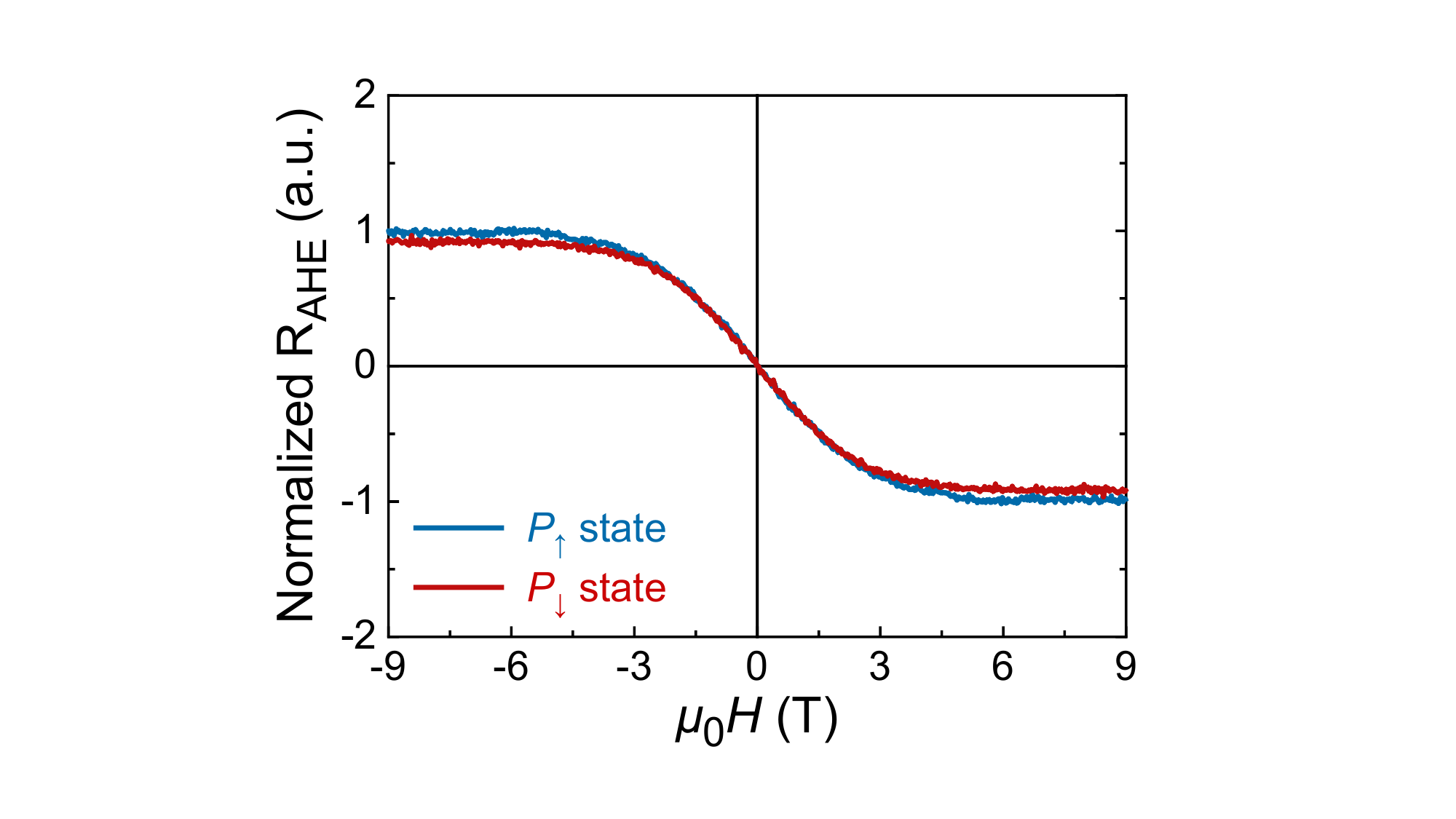}
	\caption{Normalized AHE resistance for the $P_\uparrow$ and $P_\downarrow$ states in a Pt(3 nm)/NiO(4 u.c.) heterostructure on a PMN-PT substrate.}
	\label{SF11}
\end{figure}
\end{document}